\overfullrule = 0pt

\font\bigg=cmbx10 at 17.3 truept    \font\bgg=cmbx10 at 12 truept
\font\twelverm=cmr10 scaled 1200    \font\twelvei=cmmi10 scaled 1200
\font\twelvesy=cmsy10 scaled 1200   \font\twelveex=cmex10 scaled 1200
\font\twelvebf=cmbx10 scaled 1200   \font\twelvesl=cmsl10 scaled 1200
\font\twelvett=cmtt10 scaled 1200   \font\twelveit=cmti10 scaled 1200
\def\twelvepoint{\normalbaselineskip=12.4pt
  \abovedisplayskip 12.4pt plus 3pt minus 9pt
  \belowdisplayskip 12.4pt plus 3pt minus 9pt
  \abovedisplayshortskip 0pt plus 3pt
  \belowdisplayshortskip 7.2pt plus 3pt minus 4pt
  \smallskipamount=3.6pt plus1.2pt minus1.2pt
  \medskipamount=7.2pt plus2.4pt minus2.4pt
  \bigskipamount=14.4pt plus4.8pt minus4.8pt
  \def\rm{\fam0\twelverm}          \def\it{\fam\itfam\twelveit}
  \def\sl{\fam\slfam\twelvesl}     \def\bf{\fam\bffam\twelvebf}
\def\b{\fam\bffam\twelvebf}
  \def\mit{\fam 1}                 \def\cal{\fam 2}
  \def\tt{\twelvett}
  \textfont0=\twelverm   \scriptfont0=\tenrm   \scriptscriptfont0=\sevenrm
  \textfont1=\twelvei    \scriptfont1=\teni    \scriptscriptfont1=\seveni
  \textfont2=\twelvesy   \scriptfont2=\tensy   \scriptscriptfont2=\sevensy
  \textfont3=\twelveex   \scriptfont3=\twelveex  \scriptscriptfont3=\twelveex
  \textfont\itfam=\twelveit
  \textfont\slfam=\twelvesl
  \textfont\bffam=\twelvebf \scriptfont\bffam=\tenbf
  \scriptscriptfont\bffam=\sevenbf
  \normalbaselines\rm}

\def\IZ{\relax\ifmmode\mathchoice
{\hbox{\cmss Z\kern-.4em Z}}{\hbox{\cmss Z\kern-.4em Z}}
{\lower.9pt\hbox{\cmsss Z\kern-.4em Z}}
{\lower1.2pt\hbox{\cmsss Z\kern-.4em Z}}\else{\cmss Z\kern-.4em Z}\fi}

\def\oneandahalfspace{\baselineskip=\normalbaselineskip \multiply
\baselineskip by 1}

\newcount\equationnumber
\advance\equationnumber by1
\def\ifundefined#1{\expandafter\ifx\csname#1\endcsname\relax}
\def\docref#1{\ifundefined{#1} {\bf ?.?}\message{#1 not yet defined,}
\else \csname#1\endcsname \fi}
\def\autoeqnum{\def\eqlabel##1{\edef##1{\the\equationnumber}}}
\def\no{\eqno(\the\equationnumber){\global\advance\equationnumber by1}}
\newcount\citationnumber
\advance\citationnumber by1
\def\ifundefined#1{\expandafter\ifx\csname#1\endcsname\relax}
\def\cite#1{\ifundefined{#1} {\bf ?.?}\message{#1 not yet defined,}
\else \csname#1\endcsname \fi}
\def\autocite{\def\citelabel##1{\edef##1{\the\citationnumber}\global\advance\citationnumber by1}}
\def\preprintno#1{\rightline{\rm #1}}

\hsize=6.5truein
\hoffset=.1truein
\vsize=8.9truein
\voffset=.05truein
\parskip=\medskipamount
\twelvepoint           
\oneandahalfspace
\autocite
\autoeqnum 


\def\ss{\scriptscriptstyle}
\def\cl{\centerline}

\def\lra{\longrightarrow}

\vskip -48 truept

\preprintno{ICN-UNAM-97-08}
{ \hfill 25th June, 1997}
\break
\vskip 0.6truein

\cl{\bigg {EMERGENCE OF ALGORITHMIC LANGUAGE}}
\vskip 0.3truein
\cl{\bigg {IN GENETIC SYSTEMS}}
\vskip 0.3truein
\cl{\bgg O. Angeles Palacios}
\vskip\baselineskip
\cl{\it Facultad de Ingenieria, UNAM}
\cl{\it Circuito Exterior, C.U. }
\cl{\it M\'exico D.F. 04510.}
\vskip\baselineskip
\cl{\bgg H. Waelbroeck and C.R Stephens}
\vskip\baselineskip
\cl{\it Instituto de Ciencias Nucleares, UNAM,}
\cl{\it Circuito Exterior, A.Postal 70-543,}
\cl{\it M\'exico D.F. 04510.}
\vskip 1truein

\noindent{\bf Abstract:}\ \ 
In genetic systems there is a non-trivial interface between the sequence of
symbols which constitutes the chromosome, or ``genotype'', and the products
which this sequence encodes --- the ``phenotype''. This
interface can be thought of as a ``computer''. In this case the chromosome is
viewed as an algorithm and the phenotype as the result of the computation. 
In general only a small fraction of all possible sequences of symbols 
makes any sense for a given computer. The difficulty of finding 
meaningful algorithms by random mutation is known as the {\it brittleness 
problem}. In this paper we show that mutation and crossover favour the emergence
of an algorithmic language which facilitates the production of meaningful 
sequences following random mutations of the genotype. We base our conclusions 
on an analysis of the population dynamics of a variant of Kitano's 
neurogenetic model wherein the chromosome encodes the rules 
for cellular division and the phenotype is a 16-cell organism interpreted 
as a connectivity matrix for a feedforward neural network. We show that 
an algorithmic language emerges, describe this language {\it in extenso}, 
and show how it helps to solve the brittleness problem.  
\


\vfil \eject

\line{\bgg 1. Introduction \hfil}

 Darwin's proposal that evolution proceeds by random mutation and 
natural selection has been the keystone of evolution theory since 
the XIX'th century --- yet objections 
linger on. Large mutations require coordinated changes of several 
phenotypic traits, something which seems unlikely to occur at random. 
Also, the efficiency with which species adapt to changes in the 
environment has led some to suggest that there should be a mechanism for 
environmental feedback which favours useful mutations over random ones 
(Steele 1979). 

Proposals for a direct environmental feedback that would
pre-determine the mutations have been largely discarded. 
The so-called ``central dogma'' (Lewin 1995) states that information
from the environment cannot be transferred to DNA. Actually, this 
``dogma'' is not quite true, although the conclusion that there is no 
direct environment feedback probably {\it is}. 
Viruses can incorporate their own coding in the germ line cells, as inherited 
endogenous proviruses. The enzyme methylase can induce a mutation hotspot, 
which allows for indirect information transfer through the location of the 
hotspot, etc. Yet it seems difficult for information from the environment to be 
usefully transferred through such mechanisms, and this is why the central 
dogma is so well accepted. 

In this article we will show that the environment
can organise the search for new genetic solutions {\it within the context
of random mutations of the chromosome}, the essential idea being that random
mutations of the {\it genotype} produce organised mutations of the 
{\it phenotype}. The source of this organisation is a symmetry 
breaking within the gene pool induced by the action of the genetic operators.
The symmetry breaking occurs among degenerate genotypes, or ``synonyms'', that
all map to the same phenotype. We claim that this symmetry breaking
can incorporate information about the environment and facilitate the
search of new genetic solutions. According to the theory of branching processes
(Taib 1994, Garc\'\i a-Pelayo 1994) symmetry breaking would occur 
spontaneously in a finite breeding pool,
this observation being the backbone of the Neutral Theory of molecular 
evolution (Kimura 1983). However, we emphasize that the 
symmetry breaking seen here is not spontaneous but rather is 
{\it induced} by the action of the genetic operators, such as 
mutation and recombination. 

If one considers the growth of an allele over many generations 
selection forces will take into account not only
the selective advantage of this allele but also its ability to produce 
well-adapted offspring, which can themselves produce well-adapted offspring,
etc. Since mutation and recombination act differently on synonymous alleles 
the synonyms will differ in their descendence, both in the passive 
sense of genes surviving mutations, and in the active 
sense of generating new genetic solutions. Thus, the time-averaged
effective fitness function provides a selective pressure which enhances the 
production of potentially successful mutants by selecting those synonyms
that have a higher probability to generate well-adapted offspring. 
This is a highly non-trivial proposal implying an environmental feedback 
in the genetic search such that mutant phenotypes are to some extent  
tuned to the current environment. Although this suggestion brings back the
ghost of Lamarckism we stress 
that it does not contradict the central dogma. Information from the environment is
incorporated indirectly through the symmetry breaking of the gene pool, not 
at the level of a single individual.

 The simplest manifestation of synonym symmetry in the genetic code is
the codon redundancy. For most amino acids the different synonymous 
codons which represent it have different target spaces for a 
simple point mutation. This has been shown to influence the
preference for synonymous codons (or {\it codon bias}, Grantham 1980) in 
highly mutable organisms such as the HIV retrovirus (Vera and Waelbroeck 1996, 
Mora et al 1997). 
It is easy to see how different synonyms can have different mutabilities.
Consider for example the synonymous words {\it dead}
and {\it defunct}, and assume that a mutation changes a single letter. 
The word {\it dead} can mutate to {\it deed}, {\it bead},
{\it lead}, {\it deaf}, {\it dean}, {\it dear}, {\it read} or {\it deal}, 
but it is difficult to generate a meaningful word by mutating 
the word {\it defunct}. The time-averaged effective fitness
will give a selective edge to words with the ability to mutate to another 
useful form. Since what is ``useful'' is generally environment-dependent, 
this implies that the symmetry-breaking process incorporates information 
about the environment into the gene pool. 

 In order for this mechanism to usefully guide the search for new genetic
solutions in a more complex organism one must generalise the concept of 
synonym beyond the single-codon degeneracy of the genetic code. 
The chromosome does not encode directly the size and shape 
of various parts of an organism, but an {\it interpreter}, embodied by 
the biochemical processes in and among the living cells, which 
carry the genetic information from the chromosome (genotype) to
its expression as a characteristic shape or function of an organ 
(phenotype). This process has been described as ``percolation
through scales'' (Conrad 1996). The interpreter allows for 
several types of synonym. The most trivial example is the above
mentioned codon redundancy, but there are more subtle synonyms which
involve the machinery of gene regulation, secondary structures, etc.,
for which symmetry breaking can be related to the emergence of an 
{\it algorithmic language}.

 If one views the chromosome as an algorithm, and the interpreter 
as the computer which executes this algorithm, the breaking of 
synonym symmetry is related to the selection of
a language, where ``words'' or ``grammatical rules'' are selected if they facilitate 
the search for successful mutants. This will be the case if they are related to
an approximate decomposition of the optimization problem into smaller subproblems
which are not strongly coupled.
This in turn requires that the genetic interpreter be sufficiently flexible to realise the
required decomposition, and that the fitness landscape be sufficiently correlated 
to allow for the decomposition. 
An example would be Kauffman's $Nk$ landscapes for $k << N$, together with 
his model of cellular gene regulation (Kauffman 1993).

The importance of viewing the genotype as an algorithm for a solution, 
rather than the solution itself, has been discussed previously in
the context of GAs (Asselmeyer, Ebeling and Ros\'e 1995, Adami 1994), as 
has the idea that ``intelligence'' is an emerging collective property 
(e.g., Rauch {\it et al.} 1995). Our aim in this paper is to show that
these ideas are closely related. The existence of synonyms and the
related symmetry breaking phenomenon are the 
key to understanding how intelligence emerges in the genetic search. 

 In this paper we will demonstrate the emergence of an algorithmic language in the 
context of learning in a neurogenetic model. The method we use to encode the 
neural net (NN) will be via an indirect encoding. 
More powerful interpreters have been proposed, including Kitano's original 
proposal and extensions thereof (Gruau 1992, Happel 1994, Maniezzo 1994). 
However, since our purpose here 
is only to understand the phenomenon of emergence of an algorithmic language we 
have tried to simplify the interpreter as much as possible to promote clarity.
A typical direct encoding would assign one gene for each connection. 
In contrast, indirect encoding in the present case requires that an 
interpreter generates networks through the application of a ``set of rules''. 

 Kitano compared the two methods and described advantages of his 
own indirect encoding model (Kitano 1990, 1994). The rules of the 
interpreter used in our paper are inspired from this model and will 
be described in section 3. We will  
demonstrate that the existence of such an interpreter results in the 
emergence of an adaptive property in the genetic configuration of the 
population. The property observed here during different evolution stages 
is called a language because it is possible to describe it with a set of 
rules. The effects of these rules are not present at the beginning of the 
evolution process, hence the phrase ``emergence of a language". 
The latter will be manifest in 
the apprearance of two properties: first, symmetry breaking 
within the space of genotypes, 
i.e. the preference for certain synonyms. This will be demonstrated via
an analysis of the distribution frequency of the 
configurations of different parts of the chromosome. Second, we will
show that the symmetry breaking naturally leads 
one to establish a set of rules, thus defining the 
algorithmic language, which if respected by the parent strings, 
clearly enhance the probability of a fit offspring.  

\

\line{\bgg 2. Symmetry Breaking and Indirect versus Direct Codification \hfil}

In this section we will try to illustrate the principle ideas of the paper in a
simple context and discuss the difference between direct and indirect 
codification.

Consider the different classes of maps that may be defined: 
first, $f_G:G\lra R^+$, where $G$ denotes 
the space of genotypes and $f_{\ss G}$ is the fitness function that assigns a number
to a given genotype; second, $f_{\ss Q}:\lra R^+$, where 
$Q$  is the space of phenotypes. It should be emphasized that these mappings 
may be explicitly time dependent. In fact this will normally be the case 
when the ``environment'' is time dependent. The maps may be injective or surjective. If
they are non-injective then there exist ``synonomous'' genotypes or phenotypes, i.e.
there is ``redundancy'' in the mapping. If we assume there exists a map 
$\phi:G \lra Q$ between genotype and phenotype then we have $f_{\ss G}=f_{\ss Q}\circ\phi$,
i.e. the composite map induces a fitness function on the space of genotypes (Figure 1). 
The map $\phi$ we may fruitfully think of as being an ``interpreter'', in that 
the map translates the genotypic information into something we call the phenotype,
where generically the fitness function will have a more intuitive interpretation.
As a trivial example: if one animal is faster than another it is quite easy to
understand how this enhances its fitness, however, trying to interpret this
fact at the genotypic level would be extraordinarily difficult. 
  
The question now is: what constitutes an indirect and what a direct
codification? Clearly $f_{\ss G}$ represents a ``direct'' map from genotype to $R^+$, 
whilst $f_{\ss Q}$ is a direct ``map'' from $Q$ to $R^+$, however, the composite map
$f_{\ss Q}\circ\phi$ is an ``indirect'' map. This composite map in its turn 
though is simply $f_{\ss G}$.
 The obvious point is that any indirect composite map can be written as a direct map.
The question then of a distinction between a direct codification and an indirect
codification becomes more a question of the utility of an interpreter. 

One of the principle reasons for using an interpreter, for instance in GAs, 
is that, given a set of genetic operators, the original genotypic coding may not be the 
most efficient. This is the case for a binary coding in the standard scenario where
selection, mutation and simple crossover are the preferred operators. It has been 
found that a Gray coding (Wright 1991) that maps Euclidean neighbourhoods into
Hamming neighbourhoods is more effective (Caruna and Schaffer 1988).  
The reason for the improvement is simple, in terms of the genetic operators a
Hamming metric is more natural than an Euclidean metric associated with the
landscape itself. This notion is irrespective of whether $\phi$ is an 
injective map or not. 

Here we wish to emphasize another important role 
that an interpreter may play: as an ``interpolator''. What do we mean by this? 
In the above case of faster animals the property of being faster may well be associated
with, for instance, longer legs. This is very clearly a ``macroscopic'' characteristic
of the animal. The genetic structure of its chromosomes however is very clearly a 
``microscopic'' property. The relation between the microscopic and the macroscopic in
this case is exceptionally complicated. An interpreter that maps between genotype and 
phenotype interpolates between the microscopic scale and 
the macroscopic scale. In this process however 
we pass through a whole range of intermediate scales. A successful interpreter will 
thus decompose the problem into subproblems at a much smaller scale than that of the animal.
To be simultaneously solvable these subproblems must be weakly coupled. The solutions
of these subproblems are then added together to give solutions to problems at a bigger scale.
In this sense there may well exist a hierarchy of interpreters when the macroscopic and
microscopic are very far apart. The role of a good interpreter is then to identify the appropriate
effective degrees of freedom that the system is using in passing from one scale to another.
These ``effective degrees of freedom'' constitute the elements, or words, of what
we have previously referred to as an {\it algorithmic language}.
 
We will now discuss in the context of a very simple model the phenomenon of 
induced symmetry breaking. 
The model we consider consists of four possible genotypes, $A \to D$, where
each genotype can mutate to the two adjacent genotypes when the letters
$A$, $B$, $C$, $D$ are placed clockwise on a circle. For example, $A$ can mutate
to $B$ or $D$ but not to $C$. $A$, $B$ and $D$ are synonyms in that they all encode the same
phenotype $a$, i.e. $\phi(A)=\phi(B)=\phi(D)=a$, whilst $C$ encodes the phenotype 
$c$. In a random population, $p(A) = \cdots = p(D) = {1 \over 4}$ and the 
phenotype distribution is $p(a) = {3 \over 4}$, $p(c) = {1 \over 4}$. 
If there is uniform probability $\mu$ for each genotype to mutate to an adjacent
one then the evolution equation that describes this system in the large population
limit is
$$P_i(t+1)=(1-2\mu)P'_i(t)+\mu(P'_{i-1}(t)+P'_{i+1}(t))\no$$
where $P'_i(t)=(f_i/{\bar f})P_i(t)$, $P_i(t)$ being the population fraction of genotype $i$, 
and $\bar f$ is the average fitness in the population. We assume a simple fitness 
landscape: $f_A=f_B=f_D=2$, $f_C=1$. For $\mu=0$ the steady state
population is $P(A)=P(B)=P(D)=1/3$, $P(C)=0$. Thus we see the synonym symmetry
is unbroken. However, for $\mu>0$, the genotype distribution at $t=1$ starting from a 
random distribution at $t=0$ is $P(A)=2/7$, $P(B)=P(D)=(2-\mu)/7$, $P(C)=(1+2\mu)/7$.
The corresponding phenotype distribution is: $P(a)=(6-2\mu)/7$, $P(c)=(1+2\mu)/7$.
Thus we see that there is an induced breaking of the synonym symmetry due to the
effects of mutation. 

The effective fitness (Stephens and Waelbroeck 1996) defined via
$$P_i(t+1)={f_{\ss eff_i}(t)\over {\bar f}(t)}P_i(t)\no$$
is explicitly in this model
$$f_{\ss eff_i}(t) = f_i+{\mu\over P_i(t)}(f_{i-1}P_{i-1}(t)+f_{i+1}P_{i+1}(t)-2f_iP_i(t))\no$$
At $t=0$, $f_{\ss eff_A}(0)=2$, $f_{\ss eff_B}(0)=f_{\ss eff_D}(0)=2-\mu$ and
$f_{\ss eff_C}(0)=1+2\mu$. Thus as mentioned in the introduction we see that the 
effective fitness function provides a selective pressure by selecting among the 
synonyms those that have a higher probability to produce fit descendents.  
Having introduced the principal concepts of the paper we will in the next section 
consider a more non-trivial example, a neurogenetic model, and look for the emergence of an
algorithmic language there, as a more sophisticated form of induced symmetry breaking.

\ 

\line{\bgg 3. The Neurogenetic Model \hfil}

In this section we will consider an artificial ecological environment which 
consists of a single species composed of neural networks as individuals. 
Every chromosome, or genotype, is used to produce a particular architecture 
for a feedforward NN that consists of $12$ input neurons, $4$ hidden and $1$
output neuron --- the phenotype. A GA is then applied  
to the chromosomes present in the population at each epoch. 
Thus, the GA will search 
the connectivity matrix space determined by the structure of the NN. 
Environmental effects are included in the fitness function that 
measures the learning capacity of a particular individual. 

The codification we use, as already mentioned, is indirect. A chromosome
is composed of genes,
each one of which is a three bit structure. There are eight 
different possibilities that can be labeled with a single letter as follows:
$$\vbox{\settabs 2 \columns
\+ Bit structure & label \cr 
\+000&a\cr
\+001&b\cr  
\+010&c\cr  
\+011&d\cr 
\+100&e\cr 
\+101&f\cr
\+110&g\cr  
\+111&h\cr}$$ 
A chromosome consists of eight blocks of four genes, i.e. $8\times4\times3=96$ bits
in total. The blocks themselves are also labelled, as above, from 
{\it a} to {\it h}. The reproduction process 
always begins with block {\it a}. Thus the first four genes have a priviliged role 
as they label the cells that are going to be reproduced 
in the second step of reproduction. As an example consider the chromosome
{\it baea.dcaa.defa.becd.aaea.aafh.haec.fgaa}. The two step reproduction process
specified by this chromosome can be written 
$$a \lra \pmatrix{
b & a\cr
e & a\cr}\lra
\pmatrix{
d & c & b & a\cr
a & a & e & a\cr
a & a & b & a\cr
e & a & e & a\cr
}\leftrightarrow \pmatrix{
0 & 1 & 1 & 0 & 1 & 0 & 0 & 0 & 1 & 0 & 0 & 0\cr
0 & 0 & 0 & 0 & 0 & 0 & 1 & 0 & 0 & 0 & 0 & 0\cr
0 & 0 & 0 & 0 & 0 & 0 & 0 & 0 & 1 & 0 & 0 & 0\cr
1 & 0 & 0 & 0 & 0 & 0 & 1 & 0 & 0 & 0 & 0 & 0\cr}
 \no$$
Here we see that the first block, {\it baea}, codes for the division of the original cell 
{\it a} into four cells. The first of these cells, {\it b},  then divides into four more which form the 
upper left quadrant, {\it dcaa}, of the matrix. The second cell, {\it a}, 
maps block {\it a} of the chromosome into the upper right quadrant etc. 
Finally, one constructs the connectivity 
matrix by reading left to right, row by row. In the example at hand the 
final connectivity matrix leads to the architecture seen in Figure 2. 
Thus a $1$ specifies a connection between an input neuron
and a hidden neuron and a $0$ a lack thereof.
One can see that the interpreter map, $\phi$, in this case is surjective but not injective.
For example, in the above we can change blocks {\it c}, {\it e}, {\it f}, {\it g} and {\it h}
without changing the resulting phenotype. It is also a non-local 
function on the chromosomes since entries of block number one can
target any one of the other blocks irrespective of their distance.

To define a fitness function we measured the 
learning speed of the NNs. To do so a number of training vectors is 
presented to the NN. This constitutes a single learning attempt. 
The fitness of each individual is the ratio
$$F_i  = {A_{\ss max}- A \over A_{\ss max}}.\no$$
where $A$ is the number of attempts required to reach a predetermined 
``minimum error". If this number exceeds a pre-set maximum 
value, $A_{\ss max}$, the net is assigned fitness value zero. 
As one expects that highly connected networks 
will need more information to adjust the weights of every single 
synapse, and may fall into the trap of overlearning due to the 
excessive number of adjustable parameters, this fitness function will favour the 
less connected nets with only the required connections. The possible
applications in ``network pruning'' are obvious, but we will not pursue 
further this particular line of research here.

The learning error, or the mean 
quadratic error, associated with every input vector, $p$, is 
$$E_p={1\over 2}(y_c-y_r)^2\no$$
where $y_c$ is the correct output and $y_r$ the actual output 
of the net. The correct output is a linear function of the first three components 
of each input vector:
$$y_c={\epsilon \over 3} (x_1+x_2+x_3)  + (1-\epsilon) X$$
$\epsilon$ is a noise control parameter and $X$ is a randomly
generated number. All the experiments 
were carried out with $30\%$ of noise in the signal. The input vectors 
${\buildrel \to \over x}$ are randomly generated by picking each coordinate 
independently with a  uniform probability in the unit interval. 
With this function, it should be clear that the fitness function defined
above will favour neural networks which have at least one connection to each
of the three inputs $x_i$, $i = 1, 2, 3$. Such connections we call ``effective'' 
connections. Furthermore, to avoid redundant 
connections which are likely to confuse the learning process there will 
be a bias in favour of networks which connect {\it only} to the first three 
entries. Since we know these features of the optimum network it will be 
possible to analyse the behaviour of the solutions found by the GA. 
With these choices the best input-output map with any number $(0,1,2,3)$ 
of connections to the effective inputs is a linear map. The mean squared 
errors of these optimal linear maps give lower bounds on the error of 
NNs with the same effective connections. With zero effective 
connections the optimal linear map is one which assigns the output equal to 
$1/2$ independently of the input. For this map the mean squared error with 
respect to $y_c$ is $0.0212$. The analogous values for
linear maps with $1$, $2$ and $3$ effective inputs are $0.0165$, $0.0120$ and
$0.0075$, respectively.

The minimum error, i.e. the value of error that a network has to 
reach to stop the training period, was fixed to $0.0078$. This value was chosen
in consideration of the mean squared error values of the optimal linear maps with 2 and 3 
effective inputs, given above, to guarantee that NNs are required to have all 
three effective inputs in order to be selected. 

   Once the interpreter has built the connectivity matrix the corresponding
net is trained and the synapses adjusted with the conventional backpropagation 
method. The number of input vectors that constitute an attempt was 
$100$, which means that a net was fed with a fixed set of training vectors 
which contained 100 examples. As mentioned before the maximum number of attempts 
possible was also fixed at $100$. The prediction error which is used to determine 
when the training is complete is based on a different set of vectors, called the
``testing set''.

A GA is used to search the space of 
architectures for the network that is able to learn with the smallest 
number of attempts. Our principal interest 
here is not to find the ``best" NN, but to study the search process in terms of
induced symmetry breaking and the subsequent emergence of an algorithmic
language. For the GA a population of 220 individuals was chosen, this
number being held fixed during the evolution process.
The number of generations was fixed during the experiments, 
each generation consisting of the following sequence of functions:--
training and fitness evaluation; selection; crossover and mutation.
The crossover probability was chosen to be $0.5$ and the crossover point
was chosen randomly among those points where a $12$ block of bits is not broken.
The mutation rate was expressed as a function of the population 
size: one fourth of all the offspring were submitted to a random mutation
of a single three-bit codon chosen at random in the 96-bit chromosome. This 
sets the mutation time scale at about 100 generations, which we determined
was sufficient to allow for positive selection to operate with small 
selective coefficients, down to the neutral drift cutoff at
$s \sim 1/P$, $P$ being the effective population size (Kimura 1983).

   Although in principle all the synonyms are eligible for reproduction, some
will be favoured over the others in the long term due to their superior
mutability properties. This happens because their mutation targets will 
prosper and eventually feed back to their precursor through 
reversed mutation. Another way to formulate
this argument, which is more suitable for the case of finite gene pools 
where neutral drift plays an important role (Kimura 1983), is by 
stressing that a chromosome 
that is more mutable is also the mutation target of a greater number of 
potential precursors, therefore it is more likely to be found 
by random mutation applied to all possible successful members of the 
previous generation (Waelbroeck 1997). We expect that a symmetry 
breaking will be induced when mutability 
differences favour some structures over others, and furthermore that this 
breaking will be revealed in the current context by means of an algorithmic language.
Having described the neurogenetic model, we will discuss the results of 
the numerical experiments in the remainder of the paper.

\

\line{\bgg 4. Experimental results \hfil}

   First we begin by analyzing the prediction capacities of 
certain specific NN's. We will then turn to the observed evolution dynamics
and produce evidence for the emergence of an algorithmic language. 
Finally we will consider the difference between direct and indirect encoding 
methods.

The three connectivity matrices we will consider are shown below $$\vbox{\settabs 1 \columns
\+hhhh\cr
\+hhhh\cr  
\+hhhh\cr  
\+hhhh\cr}\ \ \ \ \ \ \ \ \ \ \ \ \ \ \ \ 
\vbox{\settabs 1 \columns
\+haaa\cr
\+haaa\cr  
\+haaa\cr  
\+haaa\cr}\ \ \ \ \ \ \ \ \ \ \ \ \ \ \ \ \
\vbox{\settabs 1 \columns
\+gaaa\cr
\+gaaa\cr  
\+gaaa\cr  
\+gaaa\cr}
$$ 
The first represents a completely connected net, the second the ideal connectivity
matrix and the third a net with only two effective connections. 
Of course, the smaller the number of effective connections, the worse the 
net will perform. Recall that the correct output is the average of the first three 
entries of the 12-component input vectors. The other entries in each vector 
do not contribute to the output. 

In the following figures ($3a,b$ and $4a,b$) we 
show the prediction performance of these examples of NNs after 100 training 
iterations. Since the training vectors are randomly generated, figure $3a$ appears 
noisy; but the correlation between $y_r$ and $y_c$ is evident. 
This correlation becomes more evident if one orders
the $y_c$-$y_r$ pairs in ascending order with respect to 
$y_c$ as in figure $3b$. Following this same style, figures $4a$ and $4b$  
show the performance of the ideal network and the network with only two
of three effective connections. A comparison of fig. $3b$ and fig. $4a$ 
shows that there is no great difference between these two nets. However 
there is a notable difference if we compare the calculated fitness 
assigned to each net: while the ideal net has fitness equal to $0.62$, the 
completely connected net has fitness $0.00$, same as the network with 
two effective connections. Thus our choice of a fitness function based on the 
number of attempts required to reach a pre-assigned error value amplifies
greatly the competition between different nets. This amplification
results from the very low value of the derivative of the learning 
curve near the threshold at 0.0078. The danger of such an
amplification is that a certain amount of arbitrariness can also be 
amplified, in particular that which follows from the random choice of 
training vectors. Both the extreme amplification and the inherent 
randomness of selection are also present in natural genetic systems. 

   Having described some examples of possible NNs we will now turn to the 
main point of this paper, namely the emergence of an algorithmic language.

\noindent $\bullet$ {\it Evidence of an algorithmic language as an emergent property}. 
The experimental results are presented in three parts, each of which 
corroborates the main hypothesis underlying our work.  This 
hypothesis can be formulated in the following terms: 

\noindent {\it Given a population of individuals an interpreter 
that maps the chromosome $g$ into a phenotype $\phi(g)$, and a fitness 
function $F(\phi(g))$ that determines the expected number of offspring of
each individual evolving under the action
of the genetic operators selection, crossover and mutation, then

\noindent a) the gene pool, $G$, will organise in such a way that the search 
for new genetic solutions will be facilitated by the emergence of an 
algorithmic language, 

\noindent b) this algorithmic language will manifest itself through the 
choice of a privileged representative of each phenotype among the
possible synonymous chromosomes that encode it, and

\noindent c) it will have as chief consequence the stabilization of fitness 
inheritance from progenitors to offspring. }

\noindent Part a), the emergence of an algorithmic language in the 
population, means that a set of rules to generate individuals 
as fit, or fitter, than their parents will become apparent.

\noindent Part b), the condensation of the gene pool, implies that the
distribution of synonyms for each phenotype will be sharply peaked 
rather than diffuse; in information-theoretic terms this implies that
the gene pool {\it encodes information other than that necessary to specify
the phenotypes} (Grantham 1980).

\noindent Part c): This additional information plays a useful role in improving 
the fitness inheritance from progenitors to offspring, i.e. in solving 
the brittleness problem. 

 We examined the evolution of structure in the population, analyzing 
the number of individuals with similar ``genes'', i.e. similar blocks of 
12 bits in the chromosome. At the beginning of the evolution, the GA ``discovers" 
that those nets without the three minimum connections to the neurons 
affecting the net output have a very low fitness value and after a few 
generation a strategy emerges to avoid generating such fatally 
flawed offspring. Figure 5 shows the distribution frequency of configurations 
found in the first block of the chromosomes within a sample of 36 
individuals. It is possible to see that two strategies are preferred in 
this evolution process, the two strategies following from different 
choices of the first block, namely {\it baaa } and  {\it daaa}.

 Analyzing the second block ({\it b}) and the fourth block ({\it d}), one sees 
how the most frequent configurations, which are expressed by {\it baaa} and 
{\it daaa} respectively, achieve the goal of connecting to all three 
effective inputs when combined with those provided by the letter {\it b} or 
{\it d} from the first block. These strategies can be
illustrated by the following examples:
$$g_1 = baaa.caea.aaeb.eaca.aaaa.aabf.abac.acaa$$
$$g_2 = daaa.caea.baab.hace.aaaa.aabf.aeac.acaa$$
Already at this stage it is clear that the symmetry is broken and that this 
symmetry breaking enhances the likelihood of inheritance of genetic traits. 
Indeed, consider an individual with the first block 
{\it baaa}. The fourth block is not expressed and thus plays the role 
of ``non-coding DNA''. The choice of which word to put in this block is 
completely irrelevant from the point of view of the phenotype! However, 
it happens that the fourth block more often than not also encodes the required 
connections, in spite of the fact that it is not expressed. 
Thanks to this symmetry breaking, if as a result of crossover or mutation 
the first block of the offspring were {\it daaa }instead of {\it baaa }, 
it would still find the necessary connections. For example, the result
of crossing the two individuals above byy cutting between the second and
third blocks is
$$g_{12} = daaa.caea.aaeb.eaca.aaaa.aabf.abac.acaa$$
while the corresponding phenotype is
$$\pmatrix{
e & a & d & a\cr
c & a & a & a\cr
d & a & d & a\cr
a & a & a & a\cr
}$$
At generation 65 a new configuration appears (denoted {\it daba}), which 
combines the characteristics of the two previous strategies.  
$$g_3 = daba.caea.faaa.heca.aaaa.aabe.aabc.acaa$$
By combining both methods of aquiring the required effective connections, 
individuals that use this discovery are more resistant to the potentially 
destructive effect of mutation. This in turn opens the door to a possible 
exploration of possible mutant genes at either block {\it b} or block {\it d},
since only one of them is vital to the survival of the offspring. This is an 
example of how silent substitutions can open the door to eventual useful 
mutations, by drifting along the ``neutral net'' (Huynen 1996) to a point 
from which a simple point mutation can produce an important phenotypic 
improvement.

 If we examine the previous examples we can conclude that the offspring are 
successful independently of the crossover point. It is well worth 
stressing that crossover produces very similar and 
successful phenotypes; in this sense one can say that the fitness landscape
has been effectively smoothed. This shows that landscape smoothness, which 
is known to be an important property, especially in relation to the brittleness 
problem (Asselmeyer, Ebeling and Ros\'e 1995), is really an emerging property 
of evolution.  We also stress that blocks 2 and 4 have almost completely 
condensed their configurations, its variability in the population is much
smaller than for blocks that do not play a role in the inheritance of 
genetic traits.  

 With this sample of three examples, we note that although a  
block is not expressed in a phenotype, it can be ``prepared'' to help 
produce fit individuals if activated in the future.
The observation of this qualitative evidence in an early stage of the 
evolution paves the way for a quantitative measure and precise formulation 
of a set of rules that describe the dynamic behaviour of the 
gene pool.

 To prepare the ground for the description {\it in extenso} of the 
algorithmic language, let us analyze what happens further along 
in the evolution.  The next table, Figure 6, shows a sample of the population 
at generation 1000. One can see that the variability of the distinct parts of the 
chromosomes has decreased. One recognises general patterns that 
characterise or distinguish each block of the chromosome from the others.
For example, the first block in the last table has the 
configuration {\it febe} $53\%$ of the time.  If we denote an unspecified 
letter or ``wild card'' with an asterisk {\it *}, we observe that $97\%$ of 
the chromosomes in the sample have the configuration {\it fe**}, and $92\%$ have 
the configurations {\it *e*e}.

 The frequency of the cells {\it e} and {\it f} in the first block 
of the chromosomes leads us to ask what is happening in blocks 5 ({\it e}) and 6 
({\it f}). We note that the second and fourth positions of block number one encode 
non-effective connections since they describe the right half of the connectivity
matrix, which connects to the inputs 7-12. In the configuration {\it *e*e} this
is recognised by using the same gene ({\it e}) for both sets of non-effective 
positions. What we find in block {\it e} is that 50\% of the population has the 
configuration {\it a*aa}, 86\% have the configuration {\it a**a}, and 100\% 
have at least one {\it a} in this block. It means that block 5 ({\it e}) 
has specialised to create cells that produce a limited number of redundant
connections in the phenotype, in order to avoid overlearning problems. 
The level of condensation of configurations in block 5 is also quite strong.
 At the same time, the {\it f} in block one in a majority of the individuals 
leads us to look at block 6. We emphasise that connections 
promoted by {\it f} in the first position are effective connections in the 
phenotype, since the first and third position represent the left half of 
the connectivity matrix. Two of the three effective connections are 
found in block 6 ({\it f}). In 97\% of the chromosomes this block provides 
at least two of the three required connections by means of {\it d***}, where 
${\it d}=011$. The third connection is promoted by the second block ({\it b}).
In the latter we find that 97\% of the examples have the configuration 
{\it **f*} which provides the last connection required through ${\it f}=101$.

  The rules that specify the algorithmic language are summarised in Figure 7:

\noindent $\bullet$ At position {\it f} we find the word {\it d*d*} which provides the 
effective connections {\it *11}.

\noindent $\bullet$ The second block specialises in estabilishing the connections {\it 1*1} thanks
to the use of words {\it **f*}.

\noindent $\bullet$ At position {\it e} words with more {\it a}'s are favoured.

\noindent $\bullet$ Position {\it a} is the switchboard which guarantees that the blocks 2 and 6
place the required connections and a minimum number of redundant connections is 
introduced, thanks to the activation of block 5 for the right half of the 
connectivity matrix.

 It is worth noting that the GA is perfectly capable of 
understanding the hierarchy in the chromosome: Block {\it a} has a different
role from the other blocks since it describes the first cellular division.
Although at early stages of evolution the same block {\it a} is used 
also in the second cellular division, this is not a good strategy
because the dual function of this block as a switchboard {\it and} part of the
connectivity matrix makes it susceptible to errors in reproduction. In
the final linguistic system the use of {\it a} as a part of the connectivity 
matrix has almost completely disappeared. This confirms the existence of another 
emerging property of genetic systems (Dasgupta and McGregor 1992): The 
symmetry breaking in the gene pool reflects that the system has aquired the
ability to understand and exploit the hierarchy which the interpreter imposes 
on the set of genes. 

  Coordination between different and distant parts of the chromosome 
violates the building block hypothesis (Goldberg 1980), since the resulting 
grammatical rules result in a coordination of, e.g., blocks {\it a} and 
{\it f} which are far apart on the chromosome. We stress that this violation
of the block hypothesis occurs {\it spontaneously}, and not due to a ``deceptive''
landscape which could {\it require} the coordination of distant parts of the 
chromosome (Angeline {\it et al.} 1994 and references therein).
Indeed, since we are using a non-local interpreter it would be just as easy
to encode the same phenotypes with chromosomes which encode the necessary 
information in nearby blocks, e.g. {\it a, b, c}. The failure of the 
building block hypothesis was demonstrated also with 
theoretical arguments in (Stephens and Waelbroeck 1996).

 Another interesting point is that the system has no trouble solving the 
so-called ``competing conventions problem'' (Schaffer {\it et al.} 1992). 
The simultaneous handling of several distinct grammatical rules would present 
a risk from the point of view of the crossover operator, since by crossing 
chromosomes which use different rules or conventions one risks creating 
unfit offspring. Our results show that this leads to an effective selection 
against incompatible conventions and the subsequent 
uniformization of the language. In the description of the language 
above it is clear that there is {\it redundance} (different possible rules to 
achieve the same goal) but not {\it contradiction}: Competing conventions 
cannot become established precisely because of their inability to survive
crossover in the context of the given gene pool.

 The preference for non-connections to irrelevant inputs by means of the 
block {\it e} shows that it may not be necessary to generalise the fitness 
function by imposing additional conditions to ensure the continuity of the 
convergence process, as suggested in (Oliker {\it et al.} 1992). For example
redundant connections only provoke a marginal selective cost which 
is dwarfed by stochastic fluctuations in the evaluation process, yet
the genetic system is able to avoid them without any {\it ad hoc} modification 
of the fitness function to avoid high connectivity numbers. 
This appears to be due to the effective amplification due to the 
hierarchy which the interpreter imposes upon 
the genes: Each letter in the first block leads to 4 letters or
12 connections in the synaptic matrix, so the use of a letter (like {\it e})
which expresses a gene that produces mostly non-connections has a selective 
value 12 times as large as placing a ``zero'' in a single redundant position.

We now compare our results with those found by direct encoding.
In the most obvious direct encoding the phenotypic traits are represented one by one
along the chromosome in a binary notation: In this case this means that 
each bit in the chromosome corresponds to a particular connection in 
the connectivity matrix. This implies two key differences with respect
to indirect encoding, along with some other minor points. The key 
differences are that (1) the genotype-phenotype map is injective, i.e. 
there are (generally) no synonyms, and (2) the interpreter acts locally
on the chromosome: each phenotypic trait is associated to a particular 
position along the chromosome. Since the concept of synonym is absent we 
do not expect to find an algorithmic language; furthermore since the 
interpreter is a local function the ``computation'' which produces the 
phenotype is trivial; in that sense no language would even be necessary. 
 This trivialization of the role of the interpreter is convenient in the 
sense that it makes GA's simpler to design and interpret: This simplification
is implicitly assumed in most applications of GA's, and in the claim that
they are ``general purpose'' optimization methods. 

 The following table, Figure 8, shows a sample of the chromosomes at generation 1000. 
Comparing with the equivalent table for indirect encoding, Figure 7, one 
can observe that the condensation is not as pronounced at this moment of  
evolution. We speculate that this is due to the fact that one has removed one of 
the motors driving condensation, which is precisely the symmetry breaking 
induced by the action of the genetic operators to differentiate between 
synonyms on the basis of their offspring production.
 In short, the task of writing chromosomes that represent matrices with
all three of the required connections is not so hard with direct encoding, 
so the constraint this places on the structure of chromosomes is relatively
weaker and the condensation is not so pronounced. 
 
 The difference between the two methods can be appreciated when we compare 
the fitness history of both GAs as in Figure 9.
Direct encoding produces a higher average population fitness than 
indirect encoding up to generation 250. From that point on the situation
is reversed, although the difference is not very strong because we have
chosen a very simple landscape, it appears that the GA 
with indirect encoding sustains a higher average fitness from generation 250 
to 1000. The comparison of direct versus indirect encoding with regard
to optimisation efficiency is deserving of further research.

 This observation coincides with our previous results on the
emergence of an algorithmic language. At first the interpreter complicates 
the task of finding chromosomes with all three required connections, so its
performance is worse than that for direct encoding. This is essentially
the brittleness problem: a monkey typing in a programme would have a much
lower probability to produce a meaningful algorithm in a structured language 
than an unstructured one. The more sophisticated the 
interpreter, the harder it will be to write meaningful algorithms by random 
exploration. However, and this is our key point in this paper, once endowed  
with a language that allows one to write algorithms for this 
interpreter with a better success rate, the contrary happens (as programmers
know very well). The more structured interpreter is easier to work with 
because the very structure, once understood, becomes an aide in coordinating 
the different parts of the algorithm (functions, subroutines, etc.). This 
is what we claim happens {\it spontaneously} in GAs with 
indirect encoding. In our experiments, the first block of the chromosome, {\it febe}, is
the ``main'' function (adopting a standard {\it C} terminology) which calls the functions
{\it f} and {\it b} once and the function {\it e} twice. 
It is worth stressing that in the simple landscape considered here the average
fitness reflects mostly the negative impact of failed reproduction experiments,
as most individuals end up having a similar fitness value (roughly 0.6).
Thus, the difference in average fitness indicates that there are fewer 
reproduction attempts that fail to pass the parents' phenotypic traits 
to their offspring. In other words, the effect of the algorithmic language is 
mainly to improve the inheritance properties of phenotypic traits. 
 Of course we cannot claim that the language found would
persist indefinitely; more likely than not it would be subject to constant 
modifications through the process of neutral drift.

\

\line{\bgg 5. Conclusions \hfil}

 Using a neurogenetic ``toy model'' inspired from Kitano's work, we showed
that due to the symmetry breaking induced by the action of the genetic operators: 
i) an algorithmic language emerges spontaneously as a result of the 
selective pressure to produce viable offspring; and ii) this results in an
improved inheritance of phenotypic traits and a more structurally stable 
chromosome which allows exploration of possible new genetic improvements 
without falling into the brittleness problem. 

 The interpretation of these results follows from the realisation that 
the chromosome is an algorithm, the biochemical mechanisms responsible 
for expressing the genetic information playing the role of the computer 
which executes this algorithm, and the phenotype being the
result of the computation. The breaking of 
synonym symmetry is then seen to be related to the selection of
a language, where ``words'' or ``grammatical rules'' are selected if 
they facilitate the search for successful mutants. This will be the case 
if they are related to an approximate decomposition of the optimization 
problem into smaller subproblems. 

 The algorithmic language can only reflect a decomposition of the adaptation 
problem in past generations since causality does not allow the chromosomes 
to ``know'' their mutation targets until they have actually carried out the
mutation experiments. So the language will be successful in continuing to meet 
the changing demands of the environment only if the adaptation problem 
remains structurally the same, i.e. if the structural decomposition into
subproblems that was valid in the past continues to be valid in the future. 
We call this condition on the evolution of the adaptation landscape 
``structural decomposition stability''. One might speculate that mass extinctions 
are related to violations of this condition, for example the algorithmic 
language which allowed the dinosaur species to evolve to constantly 
meet new environmental challenges throughout the Mesosoic suddenly 
failed at the Cretacious-Tertiary extinction some 65 million years ago. On
the other hand, the more recent ``algorithmic language'' which coded 
for mammalian species was able to face the new challenges. 

 We believe that these ideas could be confirmed by analyzing the present
model with an evolving landscape. One would need to modify adiabatically
the input-output function with which the data vectors are generated. A mass
extinction should be detectable through the sudden drop in the average 
fitness of the offspring. This would indicate that the algorithmic 
language is not capable of producing successful algorithms in the new 
environment. In a different context it has recently been shown 
(Stephens et al 1997) that a GA with mutation and crossover probabilities
coded in the chromosomes themselves is capable of optimization in a time
dependent fitness landscape. Once again this is due to an induced 
symmetry breaking by the genetic operators.

\

\noindent {\bf Acknowledgements} We are grateful 
to the entire Complex Systems group under 
{\it NNCP} (http://luthien.nuclecu.unam.mx/$~$nncp) for maintaining a 
stimulating research atmosphere. This work was supported in part by
DGAPA-UNAM project IN105197. 

\vskip 0.1truein

{\centerline{\bf References}}

\noindent Adami C (1994) Learning and Complexity in Genetic 
Auto-Adaptive Systems. Preprint adap-org/9401002 from nlin-sys@xyz.lanl.gov

\noindent Angeline P J, Saunders G M and Pollack J B (1994) An Evolutionary 
Algorithm that Constructs Recurrent Neural Networks. IEEE Transactions
on Neural Networks 5:54-65

\noindent Asselmeyer T Ebeling W and Ros\'e H (1995) 
Smoothing Representation of Fitness Landscapes -- the Genotype-Phenotype
Map of Evolution. Preprint adap-org/9508002 from nlin-sys@xyz.lanl.gov,
submitted to Bio. Sys.

\noindent Caruna R A and Schaffer J D (1988) {\it Proceedings of the 5th Int. Conf.
on Machine Learning\/}, 153 (Morgan Kaufmann).

\noindent Conrad M (1996) Cross-scale information processing in evolution, 
development and intelligence. BioSystems {\bf 38}: 97-109

\noindent Dasgupta D and McGregor D R (1992) Designing Application-Specific
Neural Networks Using the Structured Genetic Algorithm. in: 
{\it Proceedings of COGANN-92 International Workshop on Combinations of Genetic
Algorithms and Neural Networks}, L D Whitley and J D Schaffer, Eds. Los Alamitos,
CA: IEEE Computer Society Press

\noindent Goldberg D E (1980) Genetic algorithms in search, optimization and 
machine learning. Addison-Wesley. USA. pp 412

\noindent Grantham R (1980) Workings of the Genetic Code. Trends in Biochemical 
Sciences 5:327-331

\noindent Gruau F (1992) Genetic Synthesis of Boolean Neural Networks with a 
Cell Rewriting Developmental Process. in: 
{\it Proceedings of COGANN-92 International Workshop on Combinations of Genetic
Algorithms and Neural Networks}, L D Whitley and J D Schaffer, Eds. Los Alamitos,
CA: IEEE Computer Society Press

\noindent Happel B L M and Murre J M J (1994) Design and Evolution of Modular
Neural Network Architectures. Neural Networks {\bf 7}:985-1004

\noindent Holland J H (1975) Adaptation in Natural and Artificial Systems. 
MIT Press, Cambridge, MA. 

\noindent Huynen M (1996) Exploring Phenotype Space Through Neutral Evolution. J 
Mol Evol 43: 165-169

\noindent Kauffman S (1993) The Origins of Order. Self-Organization and Selection 
in Evolution. Oxford University Press. USA

\noindent Kimura M (1983) The Neutral Theory of Molecular Evolution. Cambridge University
Press, Cambridge

\noindent Kitano H (1990) Designing Neural Networks Using Genetic Algorithms
with a Graph Generation System. Complex Syst. {\bf 4}: 461-476

\noindent Kitano H (1994) Neurogenetic Learning: an Integrated Method of Designing and 
Training Neural Networks Using Genetic Algorithms. Physica D 75: 225-238

\noindent Maniezzo V (1994) Genetic Evolution of the Topology and Weight 
Distribution of Neural Networks. IEEE Transactions on Neural Networks {\bf 5}:39-53

\noindent Mora J, Waelbroeck H and Stephens C R (1997) Symmetry Breaking and Adaptation: Evidence From a Simple Toy Model of a Viral Neutralization Epitope. 
National University of Mexico Preprint ICN-UNAM-97-10 (adap-org/0797)

\noindent Oliker S, Furst M and Maimon O (1992) A Distributed Genetic Algorithm 
for Neural Network Design and Training. Complex Systems {\bf 6} : 459-477

\noindent Rauch E M, Millonas M M  and  Chialvo D R (1995) 
Pattern Formation and Functionality in Swarm Models. 
Preprint adap-org/9507003 from nlin-sys@xyz.lanl.gov, submitted to
Physics Letters A 

\noindent Schaffer J D, Whitley L D and Eshelman L J (1992) Combinations
of Genetic Algorithms and Neural Networks: A Survey of the State of the Art. 
in: {\it Proceedings of COGANN-92 International Workshop on Combinations of Genetic
Algorithms and Neural Networks}, L D Whitley and J D Schaffer, Eds. Los Alamitos,
CA: IEEE Computer Society Press

\noindent Steele E J (1979) Somatic Selection and Adaptive Evolution.
Williams Wallace, Toronto

\noindent Stephens C R and Waelbroeck H (1996) Analysis of the Effective Degrees 
of Freedom in Genetic Algorithms. National University of Mexico Preprint 
ICN-UNAM-96-08 (adap-org/9611005), submitted to Phys. Rev. E; (1997) Effective Degrees 
of Freedom of Genetic Algorithms and the Block Hypothesis. National University 
of Mexico Preprint ICN-UNAM-97-01 (adap-org/0797),  {\it 
Proceedings of the Sixth Int. Conf. on Genetic Algorithms and their Applications},  
(Morgan Kaufman 1997).
 
\noindent Stephens C R, Garc\'\i a Olmedo I, Mora Vargas J and Waelbroeck H (1997)
Self-Adaptation in Evolving Systems. National University 
of Mexico Preprint ICN-UNAM-97-11 (adap-org/0797), submitted to Artificial Life.

\noindent Vera S and Waelbroeck H (1996) Symmetry Breaking 
and Adaptation: the Genetic Code of Retroviral {\it env} Proteins. 
National University of Mexico Preprint ICN-UNAM-96-09 (adap-org/9610001)

\noindent Waelbroeck H (1997) Codon Bias and Mutability in HIV Sequences. 
National University of Mexico Preprint ICN-UNAM-97- (adap-org/0797)

\noindent Wright A H, (1991) {\it Foundations of Genetic Algorithms\/}, 205 
(Morgan Kaufmann).

\end